\documentclass[twocolumn, secnumarabic,  floatfix, nofootinbib, nobalancelastpage]{revtex4-1}

\def\TitleOfPaper{Lieb's most useful contribution to density functional theory?}
\def\preprintlinklocation{http://dft.uci.edu + PAPER REF}

\usepackage{amsmath}
\usepackage{physics}
\usepackage{amssymb}
\usepackage{float}
\usepackage[dvipsnames]{xcolor}

\definecolor{TITLECOL}{rgb}{0.05,0.25,0.85}
\definecolor{CONTENTSCOL}{rgb}{0.1,0.2,0.7}
\definecolor{URLCOL}{rgb}{0,0.52,0.83}
\definecolor{LINKCOL}{rgb}{0.05,0.5,0}
\definecolor{CITECOL}{rgb}{0.25,0,0.48}
\definecolor{SECOL}{rgb}{0.07,0.31,0.80}
\definecolor{SSECOL}{rgb}{0.26,0.19,0.75}

\newcommand{\coloredtitle}[1]{\title{\textcolor{TITLECOL}{#1}}}
\newcommand{\coloredauthor}[1]{\author{\textcolor{CITECOL}{#1}}} 

\usepackage{graphicx}

\def\PublicationsLink{http://dft.uci.edu/publications.php}
\def\preprintlink{ \href{\preprintlinklocation}{\TitleOfPaper} }
\def\preprinttext{~}

\usepackage{fancyhdr}
\makeatletter
\fancypagestyle{titlepage}
{	\lhead{\textsc{\preprinttext\ ~ \href{\PublicationsLink}{~}}}
	\chead{}
	\rhead{\preprintlink}
	\lfoot{}}
\makeatother
\def\preprintlink{ 
	\href{\preprintlinklocation}
        {
~}
	}

\definecolor{Green}{rgb}{0.016,0.627,0}
\definecolor{Plum}{rgb}{0.17,0,0.45}
\definecolor{LBlue}{rgb}{0,0.34,0.45}
\definecolor{Sepia}{rgb}{0.37,0.17,0.02}
\definecolor{BurntOrange}{rgb}{0.78,0.39,0}

\usepackage{hyperref}
\hypersetup{
    breaklinks=true,
    bookmarksopen=true,
    bookmarksnumbered=true,
    colorlinks=true,
    linkcolor=LINKCOL,
    linktocpage=true,
    citecolor=CITECOL,
    filecolor=magenta,      
    urlcolor=URLCOL,
}
\def\eps{\epsilon}

\graphicspath{{figures/}}

\newcommand{\hd}[1]{{\noindent\it\textcolor{blue}{#1}:}}

\def\bea{\begin{eqnarray}}
\def\eea{\end{eqnarray}}
\def\ben{\begin{equation}}
\def\een{\end{equation}}
\def\benu{\begin{enumerate}}
\def\enu{\end{enumerate}}

\def\bei{\begin{itemize}}
\def\eei{\end{itemize}}
\def\beit{\begin{itemize}}
\def\eit{\end{itemize}}
\def\benu{\begin{enumerate}}
\def\enu{\end{enumerate}}

\def\n{n}

\def\sss{\scriptscriptstyle\rm}

\def\1var{(\bx_1...\bx\N)}

\def\half{\frac{1}{2}}

\def\br{{\bf r}}

\def\bx{{x}}

\def\N{_{\sss N}}

\def\TF{^{\rm TF}}
\def\VW{^{\rm VW}}

\def\sph_int{ {\int d^3 r}}

\begin{document}
\sf
\coloredtitle{\TitleOfPaper}

\coloredauthor{Kieron Burke}
\email{kieron@uci.edu}
\affiliation{Departments of Physics and Astronomy and of Chemistry, 
University of California, Irvine, CA 92697,  USA}
\date{\today}

\begin{abstract}
The importance of the Lieb-Simon proof of the relative exactness of
Thomas-Fermi theory in the large-$Z$ limit
to modern density functional theory (DFT) is explored.
The principle, that there is a specific semiclassical limit in which functionals
become local, implies that there exist well-defined leading functional corrections to local
approximations that become relatively exact for
the  {\em error} in local approximations in this limit.
It is argued that this principle might be used to 
greatly improve the accuracy of the thousand or so DFT calculations that
are now published each week.
A key question is how to find the leading corrections to any local density
approximation as this limit is approached.
These corrections have been explicitly derived in ridiculously simple
model systems to ridiculously high order, yielding ridiculously accurate energies.
Much analytic work is needed to use this principle
to improve realistic calculations of molecules and solids.

{\noindent \em Submitted to book in honor of Elliott Lieb's 90th birthday,
edited by Rupert L. Frank, Ari Laptev, Mathieu Lewin, \& Robert Seiringer,
published by EMS Press.}
\end{abstract}
\maketitle

\hd{Lieb}
Elliott Lieb has made many seminal contributions to mathematical physics in
general, and to density functional theory (DFT) in particular.  Among the most
famous are his establishment of the formal basis of DFT\cite{L83}, the Levy-Lieb
constrained search formulation\cite{L79,L83}, and the Lieb-Oxford bound\cite{LO81}
and its extensions\cite{LL15,FKB14,CTDF15}.
These and many others are celebrated in the current volume.
But the main focus of this essay is an entirely
different work, one that is much less well-known to modern practitioners
of DFT.  

\hd{Lieb-Simon limit}
For many years, it was understood that the original DFT, that
of Thomas and Fermi (TF)\cite{T27,F28}, should become more accurate in
a certain semiclassical
limit.  In this limit, both $N$ and $Z$ becomes large while their ratio
remains fixed, with $N$ being the number of electrons and $Z$ being the
total charge on the nuclei.
The simplest case is the neutral atom, $N=Z$.
For this
problem, Lieb and Simon\cite{LS73,L76,L81} proved that
the relative error in the energy
of a TF calculation vanishes in the limit.   
Moreover, the single-particle probability density $\n(\br)$ approaches
that of TF in a weak sense\cite{L81}.
More generally, bond distances should
be scaled at the same time (as $Z^{1/3}$), to make this
a non-trivial limit for all
DFT calculations of atoms, molecules, and solids\cite{LS77}.

\hd{Neutral atoms}
This Lieb-Simon (LS) limit has long been known in the physics community,
and the expansion
for neutral atoms has been performed to extract three terms:
\ben
E(Z) = - c_0\, Z^{7/3} + \half Z^2 - c_2\, Z^{5/3} +....,
\label{EZasy}
\een
where $c_0 \approx 0.768745$ and $c_2 \approx 0.269900$ are
fundamental constants that 
can be easily calculated to arbitrary accuracy \cite{E88,ES85,ELCB08,S80,S81}.
TF theory produces exactly and only the first term, and the coefficient
can be calculated to arbitrary accuracy by solving the TF equation\cite{LCPB09},
consistent with the LS proof.
The second term is the Scott correction\cite{S52}, and can be deduced from the energy
of an atom of non-interacting electrons, called a Bohr atom\cite{HL95}.  The last term combines
two contributions, one from the local density approximation (LDA)
exchange (2/11)\cite{B29,D30}, and another from the gradient expansion
of the non-interacting kinetic energy (9/11).
The former coefficient was carefully studied by Schwinger\cite{S81}, and a
beautiful summary appears in Englert's book\cite{E88}.

\hd{Kohn-Sham DFT}
Modern DFT calculations use the Kohn-Sham scheme\cite{KS65}, in which
only a small fraction of the total energy, called the exchange-correlation (XC)
energy, needs to be approximated as a functional of the density.
My group and collaborators
have conjectured that
the analogous statement for the XC energy is also true, namely that the local
approximation for XC becomes relatively exact in this limit\cite{PCSB06,EB09,BCGP16}.
Many expectation values of quantum operators have useful local approximations
in the density, such as the kinetic energy, but we reserve the term {\em the}
local density approximation and acronym LDA for the XC energy.
The claim is that the percentage error in LDA for XC vanishes as the LS limit is reached
(not just the relative error in the total energy).
In fact, this was 'proven' for exchange by Conlon in 1983\cite{C83}, but the proof
required a rounding of the Coulomb singularity.  The Coulomb case was finally completed by Fefferman
and Seco \cite{FS90}.
In fact, both
X alone and C alone have this property.   This is important because it has
allowed study of the leading corrections to LDA, and comparisons
made with modern generalized gradient approximations (GGAs).
Insights derived from these studies have informed some of the most recent
non-empirical functional approximations, such as PBEsol\cite{PRCV08} and SCAN\cite{SRP15}.

\hd{Ionization potential}
The LS limit concerns total energies of systems, but
interesting properties depend only on energy differences.  
A simple example of an energy difference is the ionization potential of an atom, being
the difference in energy between the neutral atom and that with a single electron removed.
This energy difference has the advantage that calculations need only be performed on spherical potentials.
We performed Hartree-Fock (HF) calculations for up to 3000 electrons, and compared with KS-DFT
using LDA for X alone.  We found that the dependence of the ionization potential across a row
persisted even as $Z\to\infty$, but that nonetheless, within numerical accuracy, the LDA-X
calculations gave the exact result in the limit\cite{CSPB11}.  
This is consistent with
the proof of Solovej for Hartree-Fock (HF) theory that the ionization potential would remain finite in this
limit\cite{S03}.  At the exchange level, LDA and PBE\cite{PBE96} converge (within the limits of numerics)
to the HF value (PBE is a particular popular GGA that reduces to LDA for a uniform
gas).  For XC, there is a small correction relative to HF, but PBE agrees with LDA, suggesting
that even for the full Schr\"odinger equation, LDA is 
becoming relatively exact for 
the ionizaton potential in the LS limit.    Moreover, a simple extended TF calculation
from Englert's book\cite{E88}, with a mild extension, appears to yield the correct limit of the
ionization potential when averaged over rows of the periodic table.

\hd{Leading corrections}
These are all numerical illustrations of conjectures.   An obvious, vital
question is: Can we derive these leading corrections analytically, even in such simple
cases as spherical systems?   For the present, the answer is no for XC.
We have merely reversed-engineered some of their properties from simple atomic calculations\cite{EB09,CCKB18}.
The next question then becomes: Can we derive the leading corrections under
any conditions?   The answer is yes, for the energy of non-interacting systems
in one-dimension.  By this, we mean making the equivalent of the TF
approximation for the KS kinetic energy, and finding the leading corrections to this
approximation in the LS limit.
The answer also shows the underlying physics, explaining where and why these
approximations become very accurate.

\hd{Notation}
The rest of this article is devoted to 1D quantum systems with (mostly) smooth
potentials, to which the LS limit also applies\cite{FLS18}.  We show that, in such cases, we
{\em can} derive the leading corrections to local density approximations.
We use Hartree atomic units, with $\hbar=m=1$, so that energies are in Hartree and
distances in Bohr radii.  Many more details and references to the original works
can be found in a recent book chapter\cite{OB21}.

We denote the eigenvalues as $\eps(j)$, with $j$ beginning at zero, and the sum
of the first $N$ eigenvalues as
\ben
E_N = \sum_{j=0}^{N-1} \eps(j).
\label{EN}
\een
Such a system can be thought of as 1D Kohn-Sham system where the potential would
be the KS potential.  The TF approximation is simply a local density approximation
to the 1D kinetic energy of same spin electrons:
\ben
T\TF[\n] = \frac{\pi^2}{6}\int dx\, n^3(x),
\een
and the total energy must be minimized keeping the particle number $N$ fixed.

\hd{WKB approximation}
First we note that, in 1D, there exists the WKB approximation\cite{W26,K26,B26} which, in its simplest
form\cite{Gb05}, states that
\ben
\int_{-\infty}^\infty dx\, p(\eps,x) = (j+\nu)\pi,~~~~~j=0,1,2...
\een
where $p(\eps,x)=\sqrt{2(\eps-v(x))}$ is the classical momentum with energy $\eps$ at position $x$,
taken to be zero if $v(x) > \eps$.  Here $\nu$ is the Maslov index\cite{MF01}, which is 1/2 for two 
turning points.   We denote solutions as $\eps^{(0)}(j)$, the WKB eigenvalues.
This formula yields the exact eigenvalues in the special cases of a particle in a box ($\nu=1$ due to
two infinite barriers) and the harmonic oscillator ($\nu=1/2$).

\hd{Simple example}
We use the Poschl-Teller (PT) well as a simple example, $v(x)=-D/\cosh^2(x)$, where
$D$ is the well-depth.   The exact eigenvalues are
\ben
\eps(j) = - (\lambda-j)^2/2,
\een
where $\lambda = \sqrt{2D +1/4}-1/2$, and $j \leq \lambda$.
The WKB eigenvalues come out to be
\ben
\eps^{(0)}(j) = - (\lambda_0-j)^2/2,
\label{WKB}
\een
where $\lambda_0 = \sqrt{2D}-1/2$.  These are usually quite accurate, but not exact.

\hd{Semiclassical limit}
Now we consider the semiclassical limit, which can be stated in several different related ways.
The simplest way is to allow $\hbar\to 0$, but keep the chemical potential fixed.   
(For a spherical system, this produces $Z\to\infty$, keeping $N/Z$ fixed.  Neutrals are the
special case of $\mu=0$.)
Thus the number of occupied
levels grows in this limit.  In terms of our parameters in 1D, this is equivalent to 
an expansion in $1/{\sqrt{D}}$ keeping $j/\lambda$ fixed, so that the WKB approximation
becomes relatively exact for any individual eigenvalue.
The WKB series can be calculated for a given potential.  It is an expansion in even powers of $\hbar$
and higher orders contain higher gradients of the potential.   For the PT well, the leading
correction to Eq. (\ref{WKB}) is of order $1/{\sqrt{D}}$.

\hd{Zero-order}
More importantly, for our purposes, is to consider the sum of the lowest $N$ occupied
levels.  This is the total energy of $N$ same-spin fermions sitting in this well.
As $\hbar\to 0$, the spacing between levels becomes negligible, and this can be
gotten by a simple integral over occupied WKB eigenvalues:
\ben
E^{(0)}_N = \int_0^{N-1} dj\, \eps^{(0)}(j)
\label{EN0}
\een
Now, not only does this approximate answer become relatively exact in the semiclassical
limit, but it is precisely the same as what the (1D non-interacting) TF approximation yields.
This result was known at least as far back as 1956\cite{MP56}.

\hd{Gradient expansion}
However, so far, all we have shown is consistency with the LS proof, in a far simpler
case than the one of interest, i.e., we have shown that TF becomes relatively exact in this limit.
Our goal is to find the leading correction to the approximation of Eq. (\ref{EN0}),
as this yields the leading corrections to TF, the local density approximation to $T$, in this case.
There is one case in which this can be (relatively) easily obtained, which is that of a slowly
varying gas, as discussed by Hohenberg and Kohn\cite{HK64} and others before them.   In that case, an appropriate
semiclassical expansion yields the correct answers.  Samaj and Percus elegantly show\cite{SPb99} in 1D:
\ben
\label{SPnv}
\n(x) = \frac{p(x)}{\pi} \left[ 1 + \frac{{v}''(x)}{12 p^4(x)} + ...\right],
\een
where $p(x)$ is evaluated at $\mu$, the Fermi energy, determined by normalizing the density to $N$ particles,
and dashes denote spatial derivatives.
Similarly, the kinetic energy density is
\ben
t(x) = \frac{p^3(x)}{2\pi} \left[ \frac{1}{3} + \frac{{v}''(x)}{4p^4(x)} + ...\right].
\label{SPtv}
\een
Inversion of Eq. (\ref{SPnv}) order-by-order in gradients of the potential and insertion in Eq. (\ref{SPtv})
yields the traditional gradient expansion
\ben
T[\n] = T\TF[\n] -\frac{1}{3} T\VW[\n] + ...,
\een
where $T\VW[\n]$
is the von Weizs\"acker kinetic energy\cite{DG90}.   This is for fully spin-polarized systems,
but is trivially related to the unpolarized case via spin-scaling\cite{OP79}
This is the 1D analog of the usual Kirzhnitz expansion\cite{Kc57} in 3D, the only difference
being in the form of the TF term and the value (and sign) of the gradient correction.

\hd{Next correction}
Returning now to finite systems,
for any value of $\hbar$, one can solve self-consistently the Euler equation
for the density of the system, and insert it into the total energy expression.
Using the TF kinetic energy yields $\n\TF(x)$, the TF approximation for the density.   This yields
\ben
E\TF = T\TF[\n\TF] + \int dx\, \n\TF(x)\, v(x).
\een
As the LS limit is approached, $E\TF$ becomes identical to $E^{(0)}_N$ of Eq. (\ref{EN0}),
exactly as required by the LS theorem.

But our interest is in the leading correction in that limit.
A crucial point is that finite systems differ qualitatively from slowly varying gases,
because they have classical turning points in 1D,  i.e., places where the classical
momentum appearing in Eq. (X) vanishes.  They divide space into a region (or regions) where $p > 0$,
called classically allowed, and the rest, which is classically forbidden (in quantum
mechanics, the latter is called evanescent, and often involves exponentially decaying wavefunctions).
In 3D, the separators is a surface, producing classically allowed regions around nuclei and forbidden
regions outside.    All molecules have forbidden regions, but
slowly varying gases (and most real
solids\cite{KCBP21}) do not.    This alters
quantitatively the leading corrections to LDA in a way that
no GGA can get right for both types of systems.

In our 1D examples, we see this explicitly
in the presence of finite
Maslov indices for one case, and zero indices for the other.
Thus, the next term in the series for $E_N$, has two distinct contributions.
The first is simply the correction due to the next term in the WKB expansion
for the eigenvalues, which depends on $v''(x)$, and gives
\ben
\Delta E^{(2a)}_N = \int_0^{N-1} dj\, \eps^{(2)}(j)
\een
This is precisely the kind of term included in the gradient expansion.
For a slowly varying gas, this will be the only contribution and the total
energy will approach that of the gradient
expansion as the variation in density is made slower.   
But for any system with turning points, there is a second contribution, due to
the error in approximating the sum by an integral.   This contribution is sensitive to the
Maslov index, and can be found using the Euler-Maclaurin summation
formula\cite{B20b}.  It has the well-known form
\ben
\Delta E^{(2b)}_N =
\half\, (\eps^{(0)}_{N-1} + \eps^{(0)}_0)
\een
Such terms vanish in the slowly-varying gas, as the sum of Eq. (\ref{EN}) is in fact an integral in the
thermodynamic limit.
This contribution is precisely the difference between the standard, end-point, Riemann sum and the
trapezoidal rule for integrals. The Riemann sum is first order accurate, but the trapezoidal
rule is second order accurate. The higher-order corrections are discussed in Ref. \cite{B20b}.

Reference \cite{B20} was the first to isolate these two contributions, and calculate them
for the Poschl-Teller well.   
Not only are the improvements very meager (and sometimes non-existent)
if we only include the first term, but the improvements are spectacular once the
right correction is included.    Errors of the order of milliHartrees are obtained,
and microHartrees if the next order is also included.

\hd{Not Poschl-Teller}
Now this 1D case does indeed provide the answer we have sought.   It is a recipe 
to extract the leading correction to the local approximation in the LS limit
that can be applied to any (sufficiently smooth) $v(x)$, even if the dependence on $v(x)$
has become quite implicit. In principle, we can convert functionals of the
potential to those of the density\cite{CLEB11,CGB13}. 
 However, our use of the PT well is a little suspect, because
in that case, the WKB series is absolutely convergent, whereas typically, we expect it to
be only asymptotic.   The next question becomes, will this work (and how well does it work)
in a case where the series is asymptotic?

\hd{Asymptotic analysis}
We answer this with asymptotic analysis, applied to another simple case, the linear well $|x|$
in the half-space $x > 0$.   Its eigenvalues are trivially related to zeroes of the Airy function.
One can easily go through the same steps as performed above for the Poschl-Teller well, but it is obviously
much more tricky to ascertain the best quality of results obtainable.   In collaboration with Michael Berry,
advanced asymptotic analysis was performed on the sum of eigenvalues \cite{BB20}.   
Unlike the PT well, there are no explicit analytic formulas for the exact values.
However, a comprehensive analysis of the asymptotic expansion to all orders takes the
place of exact results.
To understand the optimal performance acheivable, given that the series are asymptotic,
we calculated the sum of the first 10 energy levels, i.e., $E_{10}$, as accurately as we could
(using asymptotics, superasymptotics, and hyperasymptotics), to find:
\ben
E_{10} \approx 81.513\, 600\, 174\, 613\, 249\, 757\, 575\, 849\, 944\, 135\, 041\, 199
\een
to be contrasted with the exact value from Mathematica\cite{BB20}
\ben
E_{10} = 81.513\, 600\, 174\, 613\, 249\, 757\, 575\, 849\, 944\, 135\, 032\, 733
\een
This is some form of black magic.  How can an approximation be right for 33 digits, and not be exact?
We can even understand why we get 33 digits right!   The optimal truncation order is typically $\pi N$, i.e.,
the order at which additions are least.  Moreover, each order is smaller by a factor of $1/N$. 
Thus $N=10$, and $\pi N = 31$.
Finally,
hyperasymptotics reduced our errors by two more orders of magnitude.  

Of course, this tour-de-force is not designed to yield practical approximations.  
No practical DFT electronic structure calculation can match this accuracy. It is purely designed
to show that we understand the basic principles, and just how powerful the correct asymptotic expansion
can be when fully understood.   

\hd{Subdominant terms}
A further difficulty can occur when there are complex turning points for the potential.   These generate
subdominant terms (the beginning of a trans-series) that are missed by the standard WKB series.  
The simplest such example is the quartic oscillator, $x^4$.   We recently published many benchmark numbers
for this potential (generalized to include a quadratic term)\cite{OB20}.  We have also completed the asymptotic sum
analysis of the WKB series plus the leading sub-dominant corrections\cite{OB22}.   While more complicated, it is
relatively straightforward to find the sub-dominant corrections to the asymptotic sums, with
similarly spectacular results (although not quite as spectacular, with errors as large as 10$^{-21}$ for
the sum of 10 levels).

\hd{The road ahead}
So, where does this all leave us, and what is the path forward?    The crucial point is that it {\em is}
possible to derive the leading corrections to the LS limit in these simple cases.   To generalize these
results to a point where they can be practically useful, we must (a) go from one to three dimensions
(b) go from the kinetic energy to the exchange and exchange-correlation energies
(c) include Coulomb attractions to nuclei as well as Coulomb repulsion between electrons, and
(d) be able to handle multi-centered problems, not just spherical systems.
We are currently pursuing multiple avenues that take steps in these directions.  Each has its own mathematical
challenges, where help would be much appreciated.

\hd{Hard math}
Perhaps the most difficult problem is performing sums over more than a single index.   This was looked
at briefly in a very simple case in Ref. \cite{BB20}.   In that case, the number staircase for
particles that are free in two-dimensions was considered, with periodic boundary conditions (i.e., particles
on a torus).  Moreover, the relative periods in the two directions was chosen as $\sqrt 2$, making them incommensurate,
and the appearance of new levels pseudorandom.   The results found could be interpreted in two different ways.
The asymptotic sums were found to yield results as good as (and almost indistinguishable from) those
of the individual levels.   Given the difficulty of such problems, this could be regarded as a very promising
sign.   On the other hand, the results were no better than the results for individual eigenvalues, so more study
needs to be done to see if better results can be achieved.

A review of asymptotic series that appeals to physical scientists is that of Boyd\cite{B99}.
A rigorous review for mathematicians is the book of Costin\cite{C08}.   There has been
recent interest in this area from the perspective of particle physics, in the language
of solitons, resurgence, and trans-series\cite{DU14,DU17,CD17}.

\hd{Conclusions}
In conclusion, it has been established that, in very simple cases, it {\em is} possible to derive general
formulas for the leading corrections to local density approximations, and that these corrections can often
be extremely accurate.   It is conjectured that if these corrections could be found for the exchange-correlation
energy of Kohn-Sham DFT, they would yield approximate functionals of far higher accuracy than those in current use.
However, there are many challenges in generalizing these results, chief among them being the ability to sum
over more than one index and to handle degeneracies.  
On the more traditional end, a rigorous proof that XC becomes relatively exact in the LS limit would be most welcome.
Any results in this and related areas from Elliott or his many friends, preferably
before his hundredth birthday, would be most
appreciated.

\hd{Acknowledgements}
This essay is dedicated to Elliott Lieb on the occasion of his 90th birthday.  He is that rarest
of mathematical physicists, namely one whose results are quite often directly and immediately useful
to practitioners.
I thank the NSF for grant CHE-1856165 for support, and Mathieu Lewin, John Perdew,
and especially Carlos Garcia-Cerveza,
for helpful comments.

This essay is a description of physics.
Any resemblance to mathematical rigor, real or assumed, is entirely coincidental.
No theorems were violated in the writing of this manuscript (I hope).

\bibliographystyle{lucas-preprint}
\bibliography{ar}

\begin{thebibliography}{60}%
\makeatletter
\providecommand \@ifxundefined [1]{%
 \@ifx{#1\undefined}
}%
\providecommand \@ifnum [1]{%
 \ifnum #1\expandafter \@firstoftwo
 \else \expandafter \@secondoftwo
 \fi
}%
\providecommand \@ifx [1]{%
 \ifx #1\expandafter \@firstoftwo
 \else \expandafter \@secondoftwo
 \fi
}%
\providecommand \natexlab [1]{#1}%
\providecommand \enquote  [1]{``#1''}%
\providecommand \bibnamefont  [1]{#1}%
\providecommand \bibfnamefont [1]{#1}%
\providecommand \citenamefont [1]{#1}%
\providecommand \href@noop [0]{\@secondoftwo}%
\providecommand \href [0]{\begingroup \@sanitize@url \@href}%
\providecommand \@href[1]{\@@startlink{#1}\@@href}%
\providecommand \@@href[1]{\endgroup#1\@@endlink}%
\providecommand \@sanitize@url [0]{\catcode `\\12\catcode `\$12\catcode
  `\&12\catcode `\#12\catcode `\^12\catcode `\_12\catcode `\%12\relax}%
\providecommand \@@startlink[1]{}%
\providecommand \@@endlink[0]{}%
\providecommand \url  [0]{\begingroup\@sanitize@url \@url }%
\providecommand \@url [1]{\endgroup\@href {#1}{\urlprefix }}%
\providecommand \urlprefix  [0]{URL }%
\providecommand \Eprint [0]{\href }%
\@ifxundefined \urlstyle {%
  \providecommand \doi  [0]{\begingroup \@sanitize@url \@doi}%
  \providecommand \@doi [1]{\endgroup \@@startlink {\doibase
  #1}doi:\discretionary {}{}{}#1\@@endlink }%
}{%
  \providecommand \doi  [0]{doi:\discretionary{}{}{}\begingroup
  \urlstyle{rm}\Url }%
}%
\providecommand \doibase [0]{http://dx.doi.org/}%
\providecommand \Doi [0]{\begingroup \@sanitize@url \@Doi }%
\providecommand \@Doi  [1]{\endgroup\@@startlink{\doibase#1}\@@Doi}%
\providecommand \@@Doi [1]{#1\@@endlink}%
\providecommand \selectlanguage [0]{\@gobble}%
\providecommand \bibinfo  [0]{\@secondoftwo}%
\providecommand \bibfield  [0]{\@secondoftwo}%
\providecommand \translation [1]{[#1]}%
\providecommand \BibitemOpen [0]{}%
\providecommand \bibitemStop [0]{}%
\providecommand \bibitemNoStop [0]{.\EOS\space}%
\providecommand \EOS [0]{\spacefactor3000\relax}%
\providecommand \BibitemShut  [1]{\csname bibitem#1\endcsname}%
\bibitem [{\citenamefont {Lieb}(1983)}]{L83}%
  \BibitemOpen
  \bibfield  {author} {\bibinfo {author} {\bibfnamefont {E.~H.}\ \bibnamefont
  {Lieb}},\ }\href {http://dx.doi.org/10.1002/qua.560240302} {\enquote
  {\bibinfo {title} {Density functionals for coulomb systems},}\ }\bibfield
  {journal} {\bibinfo  {journal} {Int. J. Quantum Chem.},\ }\textbf {\bibinfo
  {volume} {24}},\ \bibinfo {pages} {243} (\bibinfo {year} {1983}).

\bibitem [{\citenamefont {Levy}(1979)}]{L79}%
  \BibitemOpen
  \bibfield  {author} {\bibinfo {author} {\bibfnamefont {M.}~\bibnamefont
  {Levy}},\ }\href {http://www.pnas.org/content/76/12/6062.abstract} {\enquote
  {\bibinfo {title} {Universal variational functionals of electron densities,
  first-order density matrices, and natural spin-orbitals and solution of the
  $v$-representability problem},}\ }\bibfield  {journal} {\bibinfo  {journal}
  {Proceedings of the National Academy of Sciences of the United States of
  America},\ }\textbf {\bibinfo {volume} {76}},\ \bibinfo {pages} {6062}
  (\bibinfo {year} {1979}).

\bibitem [{\citenamefont {Lieb}\ and\ \citenamefont {Oxford}(1981)}]{LO81}%
  \BibitemOpen
  \bibfield  {author} {\bibinfo {author} {\bibfnamefont {E.~H.}\ \bibnamefont
  {Lieb}}\ and\ \bibinfo {author} {\bibfnamefont {S.}~\bibnamefont {Oxford}},\
  }\href {http://dx.doi.org/10.1002/qua.560190306} {\enquote {\bibinfo {title}
  {Improved lower bound on the indirect Coulomb energy},}\ }\bibfield
  {journal} {\bibinfo  {journal} {International Journal of Quantum Chemistry},\
  }\textbf {\bibinfo {volume} {19}},\ \bibinfo {pages} {427} (\bibinfo {year}
  {1981}).

\bibitem [{\citenamefont {Lewin}\ and\ \citenamefont {Lieb}(2015)}]{LL15}%
  \BibitemOpen
  \bibfield  {author} {\bibinfo {author} {\bibfnamefont {M.}~\bibnamefont
  {Lewin}}\ and\ \bibinfo {author} {\bibfnamefont {E.~H.}\ \bibnamefont
  {Lieb}},\ }\href {https://link.aps.org/doi/10.1103/PhysRevA.91.022507}
  {\enquote {\bibinfo {title} {Improved Lieb-Oxford exchange-correlation
  inequality with a gradient correction},}\ }\bibfield  {journal} {\bibinfo
  {journal} {Phys. Rev. A},\ }\textbf {\bibinfo {volume} {91}},\ \bibinfo
  {pages} {022507} (\bibinfo {year} {2015}).

\bibitem [{\citenamefont {Feinblum}\ \emph {et~al.}(2014)\citenamefont
  {Feinblum}, \citenamefont {Kenison},\ and\ \citenamefont {Burke}}]{FKB14}%
  \BibitemOpen
  \bibfield  {author} {\bibinfo {author} {\bibfnamefont {D.~V.}\ \bibnamefont
  {Feinblum}}, \bibinfo {author} {\bibfnamefont {J.}~\bibnamefont {Kenison}},\
  and\ \bibinfo {author} {\bibfnamefont {K.}~\bibnamefont {Burke}},\ }\href
  {https://doi.org/10.1063/1.4904448} {\enquote {\bibinfo {title}
  {Communication: Testing and using the Lewin-Lieb bounds in density functional
  theory},}\ }\bibfield  {journal} {\bibinfo  {journal} {The Journal of
  Chemical Physics},\ }\textbf {\bibinfo {volume} {141}},\ \bibinfo {pages}
  {241105} (\bibinfo {year} {2014}).

\bibitem [{\citenamefont {Constantin}\ \emph {et~al.}(2015)\citenamefont
  {Constantin}, \citenamefont {Terentjevs}, \citenamefont {Della~Sala},\ and\
  \citenamefont {Fabiano}}]{CTDF15}%
  \BibitemOpen
  \bibfield  {author} {\bibinfo {author} {\bibfnamefont {L.~A.}\ \bibnamefont
  {Constantin}}, \bibinfo {author} {\bibfnamefont {A.}~\bibnamefont
  {Terentjevs}}, \bibinfo {author} {\bibfnamefont {F.}~\bibnamefont
  {Della~Sala}},\ and\ \bibinfo {author} {\bibfnamefont {E.}~\bibnamefont
  {Fabiano}},\ }\href {https://link.aps.org/doi/10.1103/PhysRevB.91.041120}
  {\enquote {\bibinfo {title} {Gradient-dependent upper bound for the
  exchange-correlation energy and application to density functional theory},}\
  }\bibfield  {journal} {\bibinfo  {journal} {Phys. Rev. B},\ }\textbf
  {\bibinfo {volume} {91}},\ \bibinfo {pages} {041120} (\bibinfo {year}
  {2015}).

\bibitem [{\citenamefont {Thomas}(1927)}]{T27}%
  \BibitemOpen
  \bibfield  {author} {\bibinfo {author} {\bibfnamefont {L.~H.}\ \bibnamefont
  {Thomas}},\ }\href {http://dx.doi.org/10.1017/S0305004100011683} {\enquote
  {\bibinfo {title} {The calculation of atomic fields},}\ }\bibfield  {journal}
  {\bibinfo  {journal} {Math. Proc. Camb. Phil. Soc.},\ }\textbf {\bibinfo
  {volume} {23}},\ \bibinfo {pages} {542} (\bibinfo {year} {1927}).

\bibitem [{\citenamefont {Fermi}(1928)}]{F28}%
  \BibitemOpen
  \bibfield  {author} {\bibinfo {author} {\bibfnamefont {E.}~\bibnamefont
  {Fermi}},\ }\href {http://dx.doi.org/10.1007/BF01351576} {\enquote {\bibinfo
  {title} {Eine statistische {M}ethode zur {B}estimmung einiger {E}igenschaften
  des {A}toms und ihre {A}nwendung auf die {T}heorie des periodischen {S}ystems
  der {E}lemente (A statistical method for the determination of some atomic
  properties and the application of this method to the theory of the periodic
  system of elements)},}\ }\bibfield  {journal} {\bibinfo  {journal}
  {Zeitschrift f\"ur Physik A Hadrons and Nuclei},\ }\textbf {\bibinfo {volume}
  {48}},\ \bibinfo {pages} {73} (\bibinfo {year} {1928}).

\bibitem [{\citenamefont {Lieb}\ and\ \citenamefont {Simon}(1973)}]{LS73}%
  \BibitemOpen
  \bibfield  {author} {\bibinfo {author} {\bibfnamefont {E.}~\bibnamefont
  {Lieb}}\ and\ \bibinfo {author} {\bibfnamefont {B.}~\bibnamefont {Simon}},\
  }\href@noop {} {\enquote {\bibinfo {title} {Thomas-Fermi Theory Revisited},}\
  }\bibfield  {journal} {\bibinfo  {journal} {Phys. Rev. Lett.},\ }\textbf
  {\bibinfo {volume} {31}},\ \bibinfo {pages} {681} (\bibinfo {year} {1973}).

\bibitem [{\citenamefont {Lieb}(1976)}]{L76}%
  \BibitemOpen
  \bibfield  {author} {\bibinfo {author} {\bibfnamefont {E.~H.}\ \bibnamefont
  {Lieb}},\ }\href@noop {} {\enquote {\bibinfo {title} {The stability of
  matter},}\ }\bibfield  {journal} {\bibinfo  {journal} {Rev. Mod. Phys.},\
  }\textbf {\bibinfo {volume} {48}},\ \bibinfo {pages} {553} (\bibinfo {year}
  {1976}).

\bibitem [{\citenamefont {Lieb}(1981)}]{L81}%
  \BibitemOpen
  \bibfield  {author} {\bibinfo {author} {\bibfnamefont {E.~H.}\ \bibnamefont
  {Lieb}},\ }\href {http://link.aps.org/doi/10.1103/RevModPhys.53.603}
  {\enquote {\bibinfo {title} {Thomas-Fermi and related theories of atoms and
  molecules},}\ }\bibfield  {journal} {\bibinfo  {journal} {Rev. Mod. Phys.},\
  }\textbf {\bibinfo {volume} {53}},\ \bibinfo {pages} {603} (\bibinfo {year}
  {1981}).

\bibitem [{\citenamefont {Lieb}\ and\ \citenamefont {Simon}(1977)}]{LS77}%
  \BibitemOpen
  \bibfield  {author} {\bibinfo {author} {\bibfnamefont {E.~H.}\ \bibnamefont
  {Lieb}}\ and\ \bibinfo {author} {\bibfnamefont {B.}~\bibnamefont {Simon}},\
  }\href@noop {} {\enquote {\bibinfo {title} {The {T}homas-{F}ermi theory of
  atoms, molecules and solids},}\ }\bibfield  {journal} {\bibinfo  {journal}
  {Advances in Mathematics},\ }\textbf {\bibinfo {volume} {23}},\ \bibinfo
  {pages} {22 } (\bibinfo {year} {1977}).

\bibitem [{\citenamefont {Englert}(1988)}]{E88}%
  \BibitemOpen
  \bibfield  {author} {\bibinfo {author} {\bibfnamefont {B.-G.}\ \bibnamefont
  {Englert}},\ }\href@noop {} {\enquote {\bibinfo {title} {Semiclassical theory
  of atoms},}\ }\bibfield  {journal} {\bibinfo  {journal} {Lec. Notes Phys.},\
  }\textbf {\bibinfo {volume} {300}} (\bibinfo {year} {1988}).

\bibitem [{\citenamefont {Englert}\ and\ \citenamefont
  {Schwinger}(1985)}]{ES85}%
  \BibitemOpen
  \bibfield  {author} {\bibinfo {author} {\bibfnamefont {B.-G.}\ \bibnamefont
  {Englert}}\ and\ \bibinfo {author} {\bibfnamefont {J.}~\bibnamefont
  {Schwinger}},\ }\href@noop {} {\enquote {\bibinfo {title} {Semiclassical
  atom},}\ }\bibfield  {journal} {\bibinfo  {journal} {Phys. Rev. A},\ }\textbf
  {\bibinfo {volume} {32}},\ \bibinfo {pages} {26} (\bibinfo {year} {1985}).

\bibitem [{\citenamefont {Elliott}\ \emph {et~al.}(2008)\citenamefont
  {Elliott}, \citenamefont {Lee}, \citenamefont {Cangi},\ and\ \citenamefont
  {Burke}}]{ELCB08}%
  \BibitemOpen
  \bibfield  {author} {\bibinfo {author} {\bibfnamefont {P.}~\bibnamefont
  {Elliott}}, \bibinfo {author} {\bibfnamefont {D.}~\bibnamefont {Lee}},
  \bibinfo {author} {\bibfnamefont {A.}~\bibnamefont {Cangi}},\ and\ \bibinfo
  {author} {\bibfnamefont {K.}~\bibnamefont {Burke}},\ }\href@noop {} {\enquote
  {\bibinfo {title} {Semiclassical Origins of Density Functionals},}\
  }\bibfield  {journal} {\bibinfo  {journal} {Phys. Rev. Lett.},\ }\textbf
  {\bibinfo {volume} {100}},\ \bibinfo {pages} {256406} (\bibinfo {year}
  {2008}).

\bibitem [{\citenamefont {Schwinger}(1980)}]{S80}%
  \BibitemOpen
  \bibfield  {author} {\bibinfo {author} {\bibfnamefont {J.}~\bibnamefont
  {Schwinger}},\ }\href@noop {} {\enquote {\bibinfo {title} {Thomas-Fermi
  model: The leading correction},}\ }\bibfield  {journal} {\bibinfo  {journal}
  {Phys. Rev. A},\ }\textbf {\bibinfo {volume} {22}},\ \bibinfo {pages} {1827}
  (\bibinfo {year} {1980}).

\bibitem [{\citenamefont {Schwinger}(1981)}]{S81}%
  \BibitemOpen
  \bibfield  {author} {\bibinfo {author} {\bibfnamefont {J.}~\bibnamefont
  {Schwinger}},\ }\href@noop {} {\enquote {\bibinfo {title} {Thomas-Fermi
  model: The second correction},}\ }\bibfield  {journal} {\bibinfo  {journal}
  {Phys. Rev. A},\ }\textbf {\bibinfo {volume} {24}},\ \bibinfo {pages} {2353}
  (\bibinfo {year} {1981}).

\bibitem [{\citenamefont {Lee}\ \emph {et~al.}(2009)\citenamefont {Lee},
  \citenamefont {Constantin}, \citenamefont {Perdew},\ and\ \citenamefont
  {Burke}}]{LCPB09}%
  \BibitemOpen
  \bibfield  {author} {\bibinfo {author} {\bibfnamefont {D.}~\bibnamefont
  {Lee}}, \bibinfo {author} {\bibfnamefont {L.~A.}\ \bibnamefont {Constantin}},
  \bibinfo {author} {\bibfnamefont {J.~P.}\ \bibnamefont {Perdew}},\ and\
  \bibinfo {author} {\bibfnamefont {K.}~\bibnamefont {Burke}},\ }\href
  {http://link.aip.org/link/?JCP/130/034107/1} {\enquote {\bibinfo {title}
  {Condition on the Kohn--Sham kinetic energy and modern parametrization of the
  Thomas--Fermi density},}\ }\bibfield  {journal} {\bibinfo  {journal} {J.
  Chem. Phys.},\ }\textbf {\bibinfo {volume} {130}},\ \bibinfo {eid} {034107}
  (\bibinfo {year} {2009}).

\bibitem [{\citenamefont {Scott}(1952)}]{S52}%
  \BibitemOpen
  \bibfield  {author} {\bibinfo {author} {\bibfnamefont {J.}~\bibnamefont
  {Scott}},\ }\href@noop {} {\enquote {\bibinfo {title} {The binding energy of
  the Thomas-Fermi atom},}\ }\bibfield  {journal} {\bibinfo  {journal} {Philos.
  Mag.},\ }\textbf {\bibinfo {volume} {43}},\ \bibinfo {pages} {859} (\bibinfo
  {year} {1952}).

\bibitem [{\citenamefont {Heilmann}\ and\ \citenamefont {Lieb}(1995)}]{HL95}%
  \BibitemOpen
  \bibfield  {author} {\bibinfo {author} {\bibfnamefont {O.~J.}\ \bibnamefont
  {Heilmann}}\ and\ \bibinfo {author} {\bibfnamefont {E.~H.}\ \bibnamefont
  {Lieb}},\ }\href {http://link.aps.org/doi/10.1103/PhysRevA.52.3628} {\enquote
  {\bibinfo {title} {Electron density near the nucleus of a large atom},}\
  }\bibfield  {journal} {\bibinfo  {journal} {Phys. Rev. A},\ }\textbf
  {\bibinfo {volume} {52}},\ \bibinfo {pages} {3628} (\bibinfo {year} {1995}).

\bibitem [{\citenamefont {Bloch}(1929)}]{B29}%
  \BibitemOpen
  \bibfield  {author} {\bibinfo {author} {\bibfnamefont {F.}~\bibnamefont
  {Bloch}},\ }\href {https://doi.org/10.1007/BF01340281} {\enquote {\bibinfo
  {title} {Bemerkung zur Elektronentheorie des Ferromagnetismus und der
  elektrischen Leitf{\"a}higkeit},}\ }\bibfield  {journal} {\bibinfo  {journal}
  {Zeitschrift f{\"u}r Physik},\ }\textbf {\bibinfo {volume} {57}},\ \bibinfo
  {pages} {545} (\bibinfo {year} {1929}).

\bibitem [{\citenamefont {Dirac}(1930)}]{D30}%
  \BibitemOpen
  \bibfield  {author} {\bibinfo {author} {\bibfnamefont {P.~A.~M.}\
  \bibnamefont {Dirac}},\ }\href {http://dx.doi.org/10.1017/S0305004100016108}
  {\enquote {\bibinfo {title} {Note on Exchange Phenomena in the {T}homas
  Atom},}\ }\bibfield  {journal} {\bibinfo  {journal} {Mathematical Proceedings
  of the Cambridge Philosophical Society},\ }\textbf {\bibinfo {volume} {26}},\
  \bibinfo {pages} {376} (\bibinfo {year} {1930}).

\bibitem [{\citenamefont {Kohn}\ and\ \citenamefont {Sham}(1965)}]{KS65}%
  \BibitemOpen
  \bibfield  {author} {\bibinfo {author} {\bibfnamefont {W.}~\bibnamefont
  {Kohn}}\ and\ \bibinfo {author} {\bibfnamefont {L.~J.}\ \bibnamefont
  {Sham}},\ }\href {http://link.aps.org/doi/10.1103/PhysRev.140.A1133}
  {\enquote {\bibinfo {title} {Self-Consistent Equations Including Exchange and
  Correlation Effects},}\ }\bibfield  {journal} {\bibinfo  {journal} {Phys.
  Rev.},\ }\textbf {\bibinfo {volume} {140}},\ \bibinfo {pages} {A1133}
  (\bibinfo {year} {1965}).

\bibitem [{\citenamefont {J.~P.~Perdew}\ and\ \citenamefont
  {Burke}(2006)}]{PCSB06}%
  \BibitemOpen
  \bibfield  {author} {\bibinfo {author} {\bibfnamefont {E.~S.}\ \bibnamefont
  {J.~P.~Perdew}, \bibfnamefont {L.~A.~Constantin}}\ and\ \bibinfo {author}
  {\bibfnamefont {K.}~\bibnamefont {Burke}},\ }\href@noop {} {\enquote
  {\bibinfo {title} {Relevance of the slowly-varying electron gas to atoms,
  molecules, and solids},}\ }\bibfield  {journal} {\bibinfo  {journal} {Phys.
  Rev. Lett.},\ }\textbf {\bibinfo {volume} {97}},\ \bibinfo {pages} {223002}
  (\bibinfo {year} {2006}).

\bibitem [{\citenamefont {Elliott}\ and\ \citenamefont {Burke}(2009)}]{EB09}%
  \BibitemOpen
  \bibfield  {author} {\bibinfo {author} {\bibfnamefont {P.}~\bibnamefont
  {Elliott}}\ and\ \bibinfo {author} {\bibfnamefont {K.}~\bibnamefont
  {Burke}},\ }\href {http://www.nrcresearchpress.com/doi/abs/10.1139/V09-095}
  {\enquote {\bibinfo {title} {Non-empirical derivation of the parameter in the
  B88 exchange functional},}\ }\bibfield  {journal} {\bibinfo  {journal}
  {Canadian Journal of Chemistry},\ }\textbf {\bibinfo {volume} {87}},\
  \bibinfo {pages} {1485} (\bibinfo {year} {2009}).

\bibitem [{\citenamefont {Burke}\ \emph {et~al.}(2016)\citenamefont {Burke},
  \citenamefont {Cancio}, \citenamefont {Gould},\ and\ \citenamefont
  {Pittalis}}]{BCGP16}%
  \BibitemOpen
  \bibfield  {author} {\bibinfo {author} {\bibfnamefont {K.}~\bibnamefont
  {Burke}}, \bibinfo {author} {\bibfnamefont {A.}~\bibnamefont {Cancio}},
  \bibinfo {author} {\bibfnamefont {T.}~\bibnamefont {Gould}},\ and\ \bibinfo
  {author} {\bibfnamefont {S.}~\bibnamefont {Pittalis}},\ }\href
  {http://scitation.aip.org/content/aip/journal/jcp/145/5/10.1063/1.4959126}
  {\enquote {\bibinfo {title} {Locality of correlation in density functional
  theory},}\ }\bibfield  {journal} {\bibinfo  {journal} {The Journal of
  Chemical Physics},\ }\textbf {\bibinfo {volume} {145}},\ \bibinfo {pages}
  {054112} (\bibinfo {year} {2016}).

\bibitem [{\citenamefont {Conlon}(1983)}]{C83}%
  \BibitemOpen
  \bibfield  {author} {\bibinfo {author} {\bibfnamefont {J.~G.}\ \bibnamefont
  {Conlon}},\ }\href {http://dx.doi.org/10.1007/BF01206884} {\enquote {\bibinfo
  {title} {Semi-classical limit theorems for Hartree-Fock theory},}\ }\bibfield
   {journal} {\bibinfo  {journal} {Communications in Mathematical Physics},\
  }\textbf {\bibinfo {volume} {88}},\ \bibinfo {pages} {133} (\bibinfo {year}
  {1983}).

\bibitem [{\citenamefont {Fefferman}(1990)}]{FS90}%
  \BibitemOpen
  \bibfield  {author} {\href@noop {} {\bibinfo {author} {\bibfnamefont {L.~A.}\
  \bibnamefont {Fefferman}, \bibfnamefont {Charles L.~Seco}}},\ }\bibfield
  {journal} {\bibinfo  {journal} {Bull. Amer. Math. Soc. (N.S.)},\ }\textbf
  {\bibinfo {volume} {23}},\ \bibinfo {pages} {525} (\bibinfo {year} {1990}).

\bibitem [{\citenamefont {Perdew}\ \emph {et~al.}(2008)\citenamefont {Perdew},
  \citenamefont {Ruzsinszky}, \citenamefont {Csonka}, \citenamefont {Vydrov},
  \citenamefont {Scuseria}, \citenamefont {Constantin}, \citenamefont {Zhou},\
  and\ \citenamefont {Burke}}]{PRCV08}%
  \BibitemOpen
  \bibfield  {author} {\bibinfo {author} {\bibfnamefont {J.~P.}\ \bibnamefont
  {Perdew}}, \bibinfo {author} {\bibfnamefont {A.}~\bibnamefont {Ruzsinszky}},
  \bibinfo {author} {\bibfnamefont {G.~I.}\ \bibnamefont {Csonka}}, \bibinfo
  {author} {\bibfnamefont {O.~A.}\ \bibnamefont {Vydrov}}, \bibinfo {author}
  {\bibfnamefont {G.~E.}\ \bibnamefont {Scuseria}}, \bibinfo {author}
  {\bibfnamefont {L.~A.}\ \bibnamefont {Constantin}}, \bibinfo {author}
  {\bibfnamefont {X.}~\bibnamefont {Zhou}},\ and\ \bibinfo {author}
  {\bibfnamefont {K.}~\bibnamefont {Burke}},\ }\href
  {http://link.aps.org/doi/10.1103/PhysRevLett.100.136406} {\enquote {\bibinfo
  {title} {Restoring the Density-Gradient Expansion for Exchange in Solids and
  Surfaces},}\ }\bibfield  {journal} {\bibinfo  {journal} {Phys. Rev. Lett.},\
  }\textbf {\bibinfo {volume} {100}},\ \bibinfo {pages} {136406} (\bibinfo
  {year} {2008}).

\bibitem [{\citenamefont {Sun}\ \emph {et~al.}(2015)\citenamefont {Sun},
  \citenamefont {Ruzsinszky},\ and\ \citenamefont {Perdew}}]{SRP15}%
  \BibitemOpen
  \bibfield  {author} {\bibinfo {author} {\bibfnamefont {J.}~\bibnamefont
  {Sun}}, \bibinfo {author} {\bibfnamefont {A.}~\bibnamefont {Ruzsinszky}},\
  and\ \bibinfo {author} {\bibfnamefont {J.~P.}\ \bibnamefont {Perdew}},\
  }\href {http://link.aps.org/doi/10.1103/PhysRevLett.115.036402} {\enquote
  {\bibinfo {title} {Strongly Constrained and Appropriately Normed Semilocal
  Density Functional},}\ }\bibfield  {journal} {\bibinfo  {journal} {Phys. Rev.
  Lett.},\ }\textbf {\bibinfo {volume} {115}},\ \bibinfo {pages} {036402}
  (\bibinfo {year} {2015}).

\bibitem [{\citenamefont {Constantin}\ \emph {et~al.}(2011)\citenamefont
  {Constantin}, \citenamefont {Snyder}, \citenamefont {Perdew},\ and\
  \citenamefont {Burke}}]{CSPB11}%
  \BibitemOpen
  \bibfield  {author} {\bibinfo {author} {\bibfnamefont {L.~A.}\ \bibnamefont
  {Constantin}}, \bibinfo {author} {\bibfnamefont {J.}~\bibnamefont {Snyder}},
  \bibinfo {author} {\bibfnamefont {J.~P.}\ \bibnamefont {Perdew}},\ and\
  \bibinfo {author} {\bibfnamefont {K.}~\bibnamefont {Burke}},\ }\href@noop {}
  {\enquote {\bibinfo {title} {Ionization potential in the limit of large
  atomic number},}\ }\bibfield  {journal} {\bibinfo  {journal} {J. Chem.
  Phys.},\ }\textbf {\bibinfo {volume} {133}},\ \bibinfo {pages} {241103}
  (\bibinfo {year} {2011}).

\bibitem [{\citenamefont {Solovej}(2003)}]{S03}%
  \BibitemOpen
  \bibfield  {author} {\bibinfo {author} {\bibfnamefont {J.}~\bibnamefont
  {Solovej}},\ }\href@noop {} {\enquote {\bibinfo {title} {The ionization
  conjecture in Hartree--Fock theory},}\ }\bibfield  {journal} {\bibinfo
  {journal} {Ann. Math.},\ }\textbf {\bibinfo {volume} {158}},\ \bibinfo
  {pages} {509} (\bibinfo {year} {2003}).

\bibitem [{\citenamefont {Perdew}\ \emph {et~al.}(1996)\citenamefont {Perdew},
  \citenamefont {Burke},\ and\ \citenamefont {Ernzerhof}}]{PBE96}%
  \BibitemOpen
  \bibfield  {author} {\bibinfo {author} {\bibfnamefont {J.~P.}\ \bibnamefont
  {Perdew}}, \bibinfo {author} {\bibfnamefont {K.}~\bibnamefont {Burke}},\ and\
  \bibinfo {author} {\bibfnamefont {M.}~\bibnamefont {Ernzerhof}},\ }\href
  {http://dx.doi.org/10.1103/PhysRevLett.77.3865} {\enquote {\bibinfo {title}
  {Generalized Gradient Approximation Made Simple},}\ }\bibfield  {journal}
  {\bibinfo  {journal} {Phys. Rev. Lett.},\ }\textbf {\bibinfo {volume} {77}},\
  \bibinfo {pages} {3865} (\bibinfo {year} {1996}),\ \bibinfo {note} {{\it
  ibid.} {\bf 78}, 1396(E) (1997)}.

\bibitem [{\citenamefont {Cancio}\ \emph {et~al.}(2018)\citenamefont {Cancio},
  \citenamefont {Chen}, \citenamefont {Krull},\ and\ \citenamefont
  {Burke}}]{CCKB18}%
  \BibitemOpen
  \bibfield  {author} {\bibinfo {author} {\bibfnamefont {A.}~\bibnamefont
  {Cancio}}, \bibinfo {author} {\bibfnamefont {G.~P.}\ \bibnamefont {Chen}},
  \bibinfo {author} {\bibfnamefont {B.~T.}\ \bibnamefont {Krull}},\ and\
  \bibinfo {author} {\bibfnamefont {K.}~\bibnamefont {Burke}},\ }\href
  {https://aip.scitation.org/doi/10.1063/1.5021597} {\enquote {\bibinfo {title}
  {Fitting a round peg into a round hole: asympotically correcting the
  generalized gradient approximation for correlation},}\ }\bibfield  {journal}
  {\bibinfo  {journal} {The Journal of Chemical Physics},\ }\textbf {\bibinfo
  {volume} {149}},\ \bibinfo {pages} {084116} (\bibinfo {year} {2018}).

\bibitem [{\citenamefont {Fournais}\ \emph {et~al.}(2018)\citenamefont
  {Fournais}, \citenamefont {Lewin},\ and\ \citenamefont {Solovej}}]{FLS18}%
  \BibitemOpen
  \bibfield  {author} {\bibinfo {author} {\bibfnamefont {S.}~\bibnamefont
  {Fournais}}, \bibinfo {author} {\bibfnamefont {M.}~\bibnamefont {Lewin}},\
  and\ \bibinfo {author} {\bibfnamefont {J.~P.}\ \bibnamefont {Solovej}},\
  }\href {https://doi.org/10.1007/s00526-018-1374-2} {\enquote {\bibinfo
  {title} {The semi-classical limit of large fermionic systems},}\ }\bibfield
  {journal} {\bibinfo  {journal} {Calculus of Variations and Partial
  Differential Equations},\ }\textbf {\bibinfo {volume} {57}},\ \bibinfo
  {pages} {105} (\bibinfo {year} {2018}).

\bibitem [{\citenamefont {Okun}\ and\ \citenamefont
  {Burke}(2022){\natexlab{a}}}]{OB21}%
  \BibitemOpen
  \bibfield  {author} {\bibinfo {author} {\bibfnamefont {P.}~\bibnamefont
  {Okun}}\ and\ \bibinfo {author} {\bibfnamefont {K.}~\bibnamefont {Burke}},\
  }\href {https://arxiv.org/abs/2105.04384} {\enquote {\bibinfo {title}
  {Semiclassics: The hidden theory behind the success of DFT},}\ }\bibfield
  {journal} {\bibinfo  {journal} {arXiv:2105.04384}} (\bibinfo {year}
  {2022}{\natexlab{a}}).

\bibitem [{\citenamefont {Wentzel}(1926)}]{W26}%
  \BibitemOpen
  \bibfield  {author} {\bibinfo {author} {\bibfnamefont {G.}~\bibnamefont
  {Wentzel}},\ }\href@noop {} {\enquote {\bibinfo {title} {Eine
  Verallgemeinerung der Quantenbedingungen f\"ur die Zwecke der
  Wellenmechanik},}\ }\bibfield  {journal} {\bibinfo  {journal} {Z. Phys.},\
  }\textbf {\bibinfo {volume} {38}},\ \bibinfo {pages} {518} (\bibinfo {year}
  {1926}).

\bibitem [{\citenamefont {Kramers}(1926)}]{K26}%
  \BibitemOpen
  \bibfield  {author} {\bibinfo {author} {\bibfnamefont {H.}~\bibnamefont
  {Kramers}},\ }\href@noop {} {\enquote {\bibinfo {title} {Wellenmechanik und
  halbz\"ahlige Quantisierung},}\ }\bibfield  {journal} {\bibinfo  {journal}
  {Z. Phys.},\ }\textbf {\bibinfo {volume} {39}},\ \bibinfo {pages} {828}
  (\bibinfo {year} {1926}).

\bibitem [{\citenamefont {Brillouin}(1926)}]{B26}%
  \BibitemOpen
  \bibfield  {author} {\bibinfo {author} {\bibfnamefont {L.}~\bibnamefont
  {Brillouin}},\ }\href@noop {} {\enquote {\bibinfo {title} {La mecanique
  ondulatoire de Schr\"odinger: une methode generale de resolution par
  approximations successives},}\ }\bibfield  {journal} {\bibinfo  {journal}
  {Compt. Rend.},\ }\textbf {\bibinfo {volume} {183}},\ \bibinfo {pages} {24}
  (\bibinfo {year} {1926}).

\bibitem [{\citenamefont {Griffiths}(2005)}]{Gb05}%
  \BibitemOpen
  \bibfield  {author} {\bibinfo {author} {\bibfnamefont {D.~J.}\ \bibnamefont
  {Griffiths}},\ }\href@noop {} {\emph {\bibinfo {title} {Introduction to
  Quantum Mechanics}}}\ (\bibinfo  {publisher} {Pearson Prentice Hall},\
  \bibinfo {address} {Upper Saddle River},\ \bibinfo {year} {2005}).

\bibitem [{\citenamefont {Maslov}\ and\ \citenamefont {Fedoriuk}(2001)}]{MF01}%
  \BibitemOpen
  \bibfield  {author} {\bibinfo {author} {\bibfnamefont {V.}~\bibnamefont
  {Maslov}}\ and\ \bibinfo {author} {\bibfnamefont {V.}~\bibnamefont
  {Fedoriuk}},\ }\href@noop {} {\emph {\bibinfo {title} {Semi-Classical
  Approximation in Quantum Mechanics}}},\ Mathematical Physics and Applied
  Mathematics\ (\bibinfo  {publisher} {Springer Netherlands},\ \bibinfo {year}
  {2001}).

\bibitem [{\citenamefont {March}\ and\ \citenamefont {Plaskett}(1956)}]{MP56}%
  \BibitemOpen
  \bibfield  {author} {\bibinfo {author} {\bibfnamefont {N.~H.}\ \bibnamefont
  {March}}\ and\ \bibinfo {author} {\bibfnamefont {J.~S.}\ \bibnamefont
  {Plaskett}},\ }\href
  {http://rspa.royalsocietypublishing.org/content/235/1202/419.abstract}
  {\enquote {\bibinfo {title} {The Relation between the
  Wentzel-Kramers-Brillouin and the Thomas-Fermi Approximations},}\ }\bibfield
  {journal} {\bibinfo  {journal} {Proceedings of the Royal Society of London.
  Series A. Mathematical and Physical Sciences},\ }\textbf {\bibinfo {volume}
  {235}},\ \bibinfo {pages} {419} (\bibinfo {year} {1956}).

\bibitem [{\citenamefont {Hohenberg}\ and\ \citenamefont {Kohn}(1964)}]{HK64}%
  \BibitemOpen
  \bibfield  {author} {\bibinfo {author} {\bibfnamefont {P.}~\bibnamefont
  {Hohenberg}}\ and\ \bibinfo {author} {\bibfnamefont {W.}~\bibnamefont
  {Kohn}},\ }\href {http://link.aps.org/doi/10.1103/PhysRev.136.B864} {\enquote
  {\bibinfo {title} {Inhomogeneous Electron Gas},}\ }\bibfield  {journal}
  {\bibinfo  {journal} {Phys. Rev.},\ }\textbf {\bibinfo {volume} {136}},\
  \bibinfo {pages} {B864} (\bibinfo {year} {1964}).

\bibitem [{\citenamefont {Samaj}\ and\ \citenamefont {Percus}(1999)}]{SPb99}%
  \BibitemOpen
  \bibfield  {author} {\bibinfo {author} {\bibfnamefont {L.}~\bibnamefont
  {Samaj}}\ and\ \bibinfo {author} {\bibfnamefont {J.~K.}\ \bibnamefont
  {Percus}},\ }\href {http://link.aip.org/link/?JCP/111/1809/1} {\enquote
  {\bibinfo {title} {Recursion representation of gradient expansion for free
  fermion ground state in one dimension},}\ }\bibfield  {journal} {\bibinfo
  {journal} {The Journal of Chemical Physics},\ }\textbf {\bibinfo {volume}
  {111}},\ \bibinfo {pages} {1809} (\bibinfo {year} {1999}).

\bibitem [{\citenamefont {Dreizler}\ and\ \citenamefont {Gross}(1990)}]{DG90}%
  \BibitemOpen
  \bibfield  {author} {\bibinfo {author} {\bibfnamefont {R.~M.}\ \bibnamefont
  {Dreizler}}\ and\ \bibinfo {author} {\bibfnamefont {E.~K.~U.}\ \bibnamefont
  {Gross}},\ }\href@noop {} {\emph {\bibinfo {title} {Density Functional
  Theory: An Approach to the Quantum Many-Body Problem}}}\ (\bibinfo
  {publisher} {Springer--Verlag},\ \bibinfo {address} {Berlin},\ \bibinfo
  {year} {1990}).

\bibitem [{\citenamefont {Oliver}\ and\ \citenamefont {Perdew}(1979)}]{OP79}%
  \BibitemOpen
  \bibfield  {author} {\bibinfo {author} {\bibfnamefont {G.}~\bibnamefont
  {Oliver}}\ and\ \bibinfo {author} {\bibfnamefont {J.}~\bibnamefont
  {Perdew}},\ }\href@noop {} {\enquote {\bibinfo {title} {Spin-density gradient
  expansion for the kinetic energy},}\ }\bibfield  {journal} {\bibinfo
  {journal} {Phys. Rev. A},\ }\textbf {\bibinfo {volume} {20}},\ \bibinfo
  {pages} {397} (\bibinfo {year} {1979}).

\bibitem [{\citenamefont {Kirzhnits}(1957)}]{Kc57}%
  \BibitemOpen
  \bibfield  {author} {\bibinfo {author} {\bibfnamefont {D.}~\bibnamefont
  {Kirzhnits}},\ }\href@noop {} {\enquote {\bibinfo {title} {Quantum
  corrections to the Thomas-Fermi equation},}\ }\bibfield  {journal} {\bibinfo
  {journal} {Sov. Phys. JETP},\ }\textbf {\bibinfo {volume} {5}},\ \bibinfo
  {pages} {64} (\bibinfo {year} {1957}).

\bibitem [{\citenamefont {Kaplan}\ \emph {et~al.}(2021)\citenamefont {Kaplan},
  \citenamefont {Clark}, \citenamefont {Burke},\ and\ \citenamefont
  {Perdew}}]{KCBP21}%
  \BibitemOpen
  \bibfield  {author} {\bibinfo {author} {\bibfnamefont {A.~D.}\ \bibnamefont
  {Kaplan}}, \bibinfo {author} {\bibfnamefont {S.~J.}\ \bibnamefont {Clark}},
  \bibinfo {author} {\bibfnamefont {K.}~\bibnamefont {Burke}},\ and\ \bibinfo
  {author} {\bibfnamefont {J.~P.}\ \bibnamefont {Perdew}},\ }\href
  {https://doi.org/10.1038/s41524-020-00479-0} {\enquote {\bibinfo {title}
  {Calculation and interpretation of classical turning surfaces in solids},}\
  }\bibfield  {journal} {\bibinfo  {journal} {npj Computational Materials},\
  }\textbf {\bibinfo {volume} {7}},\ \bibinfo {pages} {25} (\bibinfo {year}
  {2021}).

\bibitem [{\citenamefont {Burke}(2020){\natexlab{a}}}]{B20b}%
  \BibitemOpen
  \bibfield  {author} {\bibinfo {author} {\bibfnamefont {K.}~\bibnamefont
  {Burke}},\ }\href {http://dx.doi.org/10.1039/D0FD00057D} {\enquote {\bibinfo
  {title} {Deriving approximate functionals with asymptotics},}\ }\bibfield
  {journal} {\bibinfo  {journal} {Faraday Discuss.},\ }\textbf {\bibinfo
  {volume} {224}},\ \bibinfo {pages} {98} (\bibinfo {year}
  {2020}{\natexlab{a}}).

\bibitem [{\citenamefont {Burke}(2020){\natexlab{b}}}]{B20}%
  \BibitemOpen
  \bibfield  {author} {\bibinfo {author} {\bibfnamefont {K.}~\bibnamefont
  {Burke}},\ }\href {https://doi.org/10.1063/5.0002287} {\enquote {\bibinfo
  {title} {Leading correction to the local density approximation of the kinetic
  energy in one dimension},}\ }\bibfield  {journal} {\bibinfo  {journal} {The
  Journal of Chemical Physics},\ }\textbf {\bibinfo {volume} {152}},\ \bibinfo
  {pages} {081102} (\bibinfo {year} {2020}{\natexlab{b}}).

\bibitem [{\citenamefont {Cangi}\ \emph {et~al.}(2011)\citenamefont {Cangi},
  \citenamefont {Lee}, \citenamefont {Elliott}, \citenamefont {Burke},\ and\
  \citenamefont {Gross}}]{CLEB11}%
  \BibitemOpen
  \bibfield  {author} {\bibinfo {author} {\bibfnamefont {A.}~\bibnamefont
  {Cangi}}, \bibinfo {author} {\bibfnamefont {D.}~\bibnamefont {Lee}}, \bibinfo
  {author} {\bibfnamefont {P.}~\bibnamefont {Elliott}}, \bibinfo {author}
  {\bibfnamefont {K.}~\bibnamefont {Burke}},\ and\ \bibinfo {author}
  {\bibfnamefont {E.~K.~U.}\ \bibnamefont {Gross}},\ }\href
  {http://link.aps.org/doi/10.1103/PhysRevLett.106.236404} {\enquote {\bibinfo
  {title} {Electronic Structure via Potential Functional Approximations},}\
  }\bibfield  {journal} {\bibinfo  {journal} {Phys. Rev. Lett.},\ }\textbf
  {\bibinfo {volume} {106}},\ \bibinfo {pages} {236404} (\bibinfo {year}
  {2011}).

\bibitem [{\citenamefont {Cangi}\ \emph {et~al.}(2013)\citenamefont {Cangi},
  \citenamefont {Gross},\ and\ \citenamefont {Burke}}]{CGB13}%
  \BibitemOpen
  \bibfield  {author} {\bibinfo {author} {\bibfnamefont {A.}~\bibnamefont
  {Cangi}}, \bibinfo {author} {\bibfnamefont {E.~K.~U.}\ \bibnamefont
  {Gross}},\ and\ \bibinfo {author} {\bibfnamefont {K.}~\bibnamefont {Burke}},\
  }\href@noop {} {\enquote {\bibinfo {title} {Potential functionals versus
  density functionals},}\ }\bibfield  {journal} {\bibinfo  {journal} {Phys.
  Rev. A},\ }\textbf {\bibinfo {volume} {88}} (\bibinfo {year} {2013}).

\bibitem [{\citenamefont {Berry}\ and\ \citenamefont {Burke}(2020)}]{BB20}%
  \BibitemOpen
  \bibfield  {author} {\bibinfo {author} {\bibfnamefont {M.~V.}\ \bibnamefont
  {Berry}}\ and\ \bibinfo {author} {\bibfnamefont {K.}~\bibnamefont {Burke}},\
  }\href {https://doi.org/10.1088/1751-8121/ab69a6} {\enquote {\bibinfo {title}
  {Exact and approximate energy sums in potential wells},}\ }\bibfield
  {journal} {\bibinfo  {journal} {Journal of Physics A: Mathematical and
  Theoretical},\ }\textbf {\bibinfo {volume} {53}},\ \bibinfo {pages} {095203}
  (\bibinfo {year} {2020}).

\bibitem [{\citenamefont {Okun}\ and\ \citenamefont {Burke}(2021)}]{OB20}%
  \BibitemOpen
  \bibfield  {author} {\bibinfo {author} {\bibfnamefont {P.}~\bibnamefont
  {Okun}}\ and\ \bibinfo {author} {\bibfnamefont {K.}~\bibnamefont {Burke}},\
  }\href {https://onlinelibrary.wiley.com/doi/abs/10.1002/qua.26554} {\enquote
  {\bibinfo {title} {Uncommonly accurate energies for the general quartic
  oscillator},}\ }\bibfield  {journal} {\bibinfo  {journal} {International
  Journal of Quantum Chemistry},\ }\textbf {\bibinfo {volume} {121}},\ \bibinfo
  {pages} {e26554} (\bibinfo {year} {2021}).

\bibitem [{\citenamefont {Okun}\ and\ \citenamefont
  {Burke}(2022){\natexlab{b}}}]{OB22}%
  \BibitemOpen
  \bibfield  {author} {\bibinfo {author} {\bibfnamefont {P.}~\bibnamefont
  {Okun}}\ and\ \bibinfo {author} {\bibfnamefont {K.}~\bibnamefont {Burke}},\
  }\href@noop {} {\enquote {\bibinfo {title} {Asymptotics of eigenvalue sums
  when some turning points are complex},}\ }\bibfield  {journal} {\bibinfo
  {journal} {in preparation}} (\bibinfo {year} {2022}{\natexlab{b}}).

\bibitem [{\citenamefont {Boyd}(1999)}]{B99}%
  \BibitemOpen
  \bibfield  {author} {\bibinfo {author} {\bibfnamefont {J.~P.}\ \bibnamefont
  {Boyd}},\ }\href {https://doi.org/10.1023/A:1006145903624} {\enquote
  {\bibinfo {title} {The Devil's Invention: Asymptotic, Superasymptotic and
  Hyperasymptotic Series},}\ }\bibfield  {journal} {\bibinfo  {journal} {Acta
  Applicandae Mathematica},\ }\textbf {\bibinfo {volume} {56}},\ \bibinfo
  {pages} {1} (\bibinfo {year} {1999}).

\bibitem [{\citenamefont {Costin}(2008)}]{C08}%
  \BibitemOpen
  \bibfield  {author} {\bibinfo {author} {\bibfnamefont {O.}~\bibnamefont
  {Costin}},\ }\href {https://doi.org/10.1201/9781420070323} {\emph {\bibinfo
  {title} {Asymptotics and Borel Summability (1st ed)}}}\ (\bibinfo
  {publisher} {Chapman and Hall/CRC.},\ \bibinfo {year} {2008}).

\bibitem [{\citenamefont {Dunne}\ and\ \citenamefont {\"Unsal}(2014)}]{DU14}%
  \BibitemOpen
  \bibfield  {author} {\bibinfo {author} {\bibfnamefont {G.~V.}\ \bibnamefont
  {Dunne}}\ and\ \bibinfo {author} {\bibfnamefont {M.}~\bibnamefont
  {\"Unsal}},\ }\href {https://link.aps.org/doi/10.1103/PhysRevD.89.105009}
  {\enquote {\bibinfo {title} {Uniform WKB, multi-instantons, and resurgent
  trans-series},}\ }\bibfield  {journal} {\bibinfo  {journal} {Phys. Rev. D},\
  }\textbf {\bibinfo {volume} {89}},\ \bibinfo {pages} {105009} (\bibinfo
  {year} {2014}).

\bibitem [{\citenamefont {Dunne}\ and\ \citenamefont {{\"U}nsal}(2017)}]{DU17}%
  \BibitemOpen
  \bibfield  {author} {\bibinfo {author} {\bibfnamefont {G.~V.}\ \bibnamefont
  {Dunne}}\ and\ \bibinfo {author} {\bibfnamefont {M.}~\bibnamefont
  {{\"U}nsal}},\ }in\ \href@noop {} {\emph {\bibinfo {booktitle} {Resurgence,
  Physics and Numbers}}},\ \bibinfo {editor} {edited by\ \bibinfo {editor}
  {\bibfnamefont {F.}~\bibnamefont {Fauvet}}, \bibinfo {editor} {\bibfnamefont
  {D.}~\bibnamefont {Manchon}}, \bibinfo {editor} {\bibfnamefont
  {S.}~\bibnamefont {Marmi}},\ and\ \bibinfo {editor} {\bibfnamefont
  {D.}~\bibnamefont {Sauzin}}}\ (\bibinfo  {publisher} {Scuola Normale
  Superiore},\ \bibinfo {address} {Pisa},\ \bibinfo {year} {2017})\ pp.\
  \bibinfo {pages} {249--298},\ ISBN \bibinfo {isbn} {978-88-7642-613-1}.

\bibitem [{\citenamefont {Costin}\ and\ \citenamefont {Dunne}(2017)}]{CD17}%
  \BibitemOpen
  \bibfield  {author} {\bibinfo {author} {\bibfnamefont {O.}~\bibnamefont
  {Costin}}\ and\ \bibinfo {author} {\bibfnamefont {G.~V.}\ \bibnamefont
  {Dunne}},\ }\href {https://doi.org/10.1088/1751-8121/aa9e30} {\enquote
  {\bibinfo {title} {Convergence from divergence},}\ }\bibfield  {journal}
  {\bibinfo  {journal} {Journal of Physics A: Mathematical and Theoretical},\
  }\textbf {\bibinfo {volume} {51}},\ \bibinfo {pages} {04LT01} (\bibinfo
  {year} {2017}).

\end{thebibliography}%

\label{page:end}
\end{document}